\documentstyle[12pt]{article}
\begin{document}

\centerline{\bf Quantum Coherence Oscillations in Antiferromagnetic Chains }  
\baselineskip=22pt
\vspace{2cm}
\centerline{ Marius Grigorescu and Mahi R. Singh} 

\centerline{ Department of Physics and Astronomy}
\baselineskip=12pt
\centerline{ University of Western Ontario}
\centerline{ London, Ontario, Canada N6A 3K7}
\vspace{5cm}
{\bf Abstract}: \\ 
Macroscopic quantum coherence oscillations in mesoscopic antiferromagnets 
appear when the anisotropy potential creates a barrier
between the antiferromagnetic states with opposite orientations
of the Neel vector. This phenomenon is studied for the physical situation
of the nuclear spin system of eight Xe atoms arranged on a magnetic surface
along a chain. The oscillation period is calculated as a function of the
chain constant. The environmental decoherence effects at finite temperature
are accounted assuming a dipole coupling between the spin chain and
the fluctuating magnetic field of the surface. The numerical calculations
indicate that the oscillations are damped by a rate $\sim (N-1)/ \tau$,
where $N$ is the number of spins and $\tau$ is the relaxation time of
a single spin. 
\\[1cm]

\newpage

{\bf I. Introduction} \\[.5cm] \indent
The  observation of macroscopic quantum coherence (MQC) phenomena in complex
many-particle systems represents a subject of wide interest, ranging from
the  conceptual
foundations of quantum mechanics  \cite{zurek}, to the physics of the 
microelectronic devices. During the last years a particular attention was
given to the macroscopic quantum tunneling and quantum coherence
oscillations \cite{leggett}. In these phenomena 
the quantum dynamics of a prepared non-stationary wave-packet is
directly reflected by the non-classical behavior of a macroscopic
observable. However, it is difficult to identify physical
situations where such coexistence of the classical and
quantum aspects could be observed.
\\ \indent
Quantum coherence oscillations may occur in the localization probability
of relatively complex systems as the individual Xe atoms trapped in
the  surface-tip junction of the scanning tunneling microscope
\cite{mg1,mg2,mg3,tvs}. Particularly suitable candidates to observe MQC
phenomena are also the magnetic systems \cite{th}. In the anisotropic
antiferromagnets the Neel vector may change the orientation by
quantum tunneling \cite{bb} or quantum coherence oscillations
\cite{ac}. The observation of these oscillations still represents a
challenging problem, but important results have been obtained from
the measurements of the ac magnetic susceptibility in 
antiferromagnetic (AF) ferritin \cite{gad1, gad2}.
\\ \indent
The study of the non-elementary excitations in AF systems 
is also essential for understanding the high-$T_c$ superconductivity 
\cite{singh1,sudip}. Therefore, a detailed study of the MQC
phenomena in antiferromagnets appears highly interesting.
\\ \indent
The purpose of this work is to investigate the occurence of MQC oscillations
of the Neel vector in an anisotropic AF chain ${\cal C} \equiv 
\{ \vec{\bf I}_i, i=1,8 \}$ of eight spins 1/2.
The Hamiltonian correspond to the system of nuclear
spins for eight Xe atoms placed on a magnetic surface and
coupled by the magnetic dipole interaction.
Such chain structures could be constructed, for instance, 
using the scanning tunneling microscope in the "atomic switch"  
operation mode,  proved during the last years to be an efficient instrument
to displace in a controlled way the Xe atoms on a Ni surface \cite{eig}.
\\ \indent
The model Hamiltonian and the spectrum of the spin chain are presented in
Sect. II. It is found that the first two levels are quasi-degenerate, and
dominated by AF states with opposite orientations of the Neel vector.
Therefore, non-stationary AF wave packets  which are interchanged
periodically by MQC oscillations can be constructed.
Their oscillation period is calculated as a function of the
chain constant.  \\ \indent
The relaxation effects at finite temperature due to the magnetic coupling
between the spin chain and the surface are discussed in Sect. III.
The coupling between the electronic and the nuclear magnetic moments is
significant, and
in the opposite situation, when the nuclear spins play the role of the
environment, the MQC oscillations in small magnetic particles can be
suppressed \cite{garg}.  Here the environment is represented by the surface,
and the coupling is reflected by the finite relaxation time $\tau$ of the
nuclear spins.  The effects of the spin relaxation on the nuclear MQC
oscillations will be accounted by the coupling between the chain and the
fluctuating magnetic field of the magnetic surface. It is found that
when $\tau$ is in the range of seconds the MQC oscillations are damped, but
not completely suppressed. The conclusions are summarized in Sect. IV.
\\[.5cm] 
{\bf II. Quantum coherence oscillations in the AF nuclear spin chain}  \\[.5cm]
\indent
The chain considered in the present calculations
consists of $N=8$  spins 1/2 equally spaced by $d$, and interacting by the
model Hamiltonian
\begin{equation}
H_0 = J
\sum_{i=1}^{N-1} \lbrack c_x {\bf I}_{i,x} {\bf I}_{i+1,x} + c_y
{\bf I}_{i,y} {\bf I}_{i+1,y}
+ c_z {\bf I}_{i,z} {\bf I}_{i+1,z} \rbrack~~. 
\end{equation}
Here $J>0$ is the coupling strength and $c_x$, $c_y$ and $c_z$ are
anisotropy coefficients.  Classically, the AF ordering
is described by using two sub-chains, ${\cal C}_o$ and ${\cal C}_e$,
containing the odd, $\{ \vec{\bf I}_i, i=1,3,5,..., 2N-1 \}$ and 
the even spins $\{ \vec{\bf I}_i,  i=2,4,6,..., 2N \}$, respectively.
The corresponding magnetization vectors are 
$ \vec{M}_o = \gamma \hbar \sum_{i=odd} \vec{\bf I}_i$
and
$ \vec{M}_e = \gamma \hbar \sum_{i=even} \vec{\bf I}_i$, where
$\gamma$ denotes the gyromagnetic factor.  The equilibrium  configuration
of this system is antiferromagnetic when
$\vec{M}_o$ and $\vec{M}_e$ have the same magnitude $M_0$, but a relative
antiparallel orientation. The energy in this case is a function of the Neel
vector $\vec{n}= (\vec{M}_o - \vec{M}_e)/2 M_0$, and is expressed by
\begin{equation}
E^A ( \vec{n} ) = - J (N-1) (c_x n_x^2 +c_yn_y^2 +c_z n_z^2)/4~~.
\end{equation}
An anisotropic system with $c_{x,y} < c_z$ has two degenerate minima, 
$E^A_{min} = -J(N-1) c_z /4$ attained when $\vec{n}$ has the two possible 
orientations along the Z-axis, $\vec{n} = \pm \vec{e}_z$. These minima
are separated by a two-dimensional potential barrier with the maximum
$E^A_{max} ( \phi ) = - J (N-1) (c_x \cos^2 \phi +c_y \sin^2 \phi)/4$,
in the X-Y plane, where $\phi$ denotes the angle between the Neel vector
and the X-axis.
\\ \indent
A physical situation of interest described by the anisotropic Hamiltonian
of Eq. (1) appears when ${\cal C} = 
\{  \vec{ \bf I}_i, i=1,8 \}$ consists of the nuclear spins
for a chain of Xe atoms placed on a magnetic surface. Such chains
can be constructed, for instance, using techniques of atomic manipulations.
The interaction between the Xe nuclear spins is due to the magnetic dipole
forces, and the effect of the surface can be taken into account using the
simple method of images. In a coordinate system with the
Z-axis normal to the surface and the Y-axis along the chain,
each magnetic moment $\vec{m}=(m_x,m_y,mz)$ lying above the surface at the
distance $z=r_0=2.17$ \AA, equal to the radius of the Xe atom,
has an image $\vec{m}'=(-m_x, -m_y, m_z)$ at $z=-r_0$. The nearest
neighbors of $\vec{m}$ interact also with $\vec{m}'$, 
and the Hamiltonian of the whole system is expressed by Eq. (1),
where
$ J = \mu_0 \hbar^2 \gamma^2 / (4 \pi d^3)  $,
 $\mu_0 = 4 \pi 10^{-7} N/A^2$ is the vacuum permeability,
$\gamma= - 1.54 \mu_N/ \hbar $ is
the gyromagnetic factor of the $^{129}$Xe isotope and $d$ is the chain
constant. The anisotropy coefficients are
$$
c_x= 1-2 \sin^3 \alpha
$$
\begin{equation}
c_y=-2-2 \sin^3 \alpha (1-3 \sin^2 \alpha)
\end{equation}
$$
c_z= 1 + 2 \sin^3 \alpha (1-3 \cos^2 \alpha)
$$
with $\tan \alpha = d/2r_0$.
\\ \indent
The spectrum of $H_0$ was calculated for $6.8$ \AA $<d<$ $9.4$ \AA$~~$
by solving numerically the eigenvalue equation
\begin{equation}
H_0 \vert \psi_n > = E_n \vert \psi_n > ~~.
\end{equation}
The basis was defined by the common eigenstates for the Z-components of all
the spin operators ${\bf I}_{i,z}$, i=1,8. These basis states are denoted by
$\vert k> \equiv  \vert m_1^k,m_2^k,m_3^k,m_4^k,m_5^k,m_6^k,m_7^k,m_8^k>, 
~~k=1,2^8$, with $m_i^k=  \pm 1/2 $. 
The eigenstates of $H_0$ have the general form  $ \vert \psi_n> =
\sum_{k} y_k^n \vert k>$. It is found that the ground ($n=0$), and the
first excited state ($n=1$), have the largest overlap  with only
two basis states, denoted $\vert \uparrow>$ and
$\vert \downarrow>$, which are antiferromagnetic in the classical sense.
Explicitly these basis states are
\begin{equation}
\vert \uparrow>  = \vert 
\frac{1}{2}, -\frac{1}{2}, \frac{1}{2}, - \frac{1}{2}, 
\frac{1}{2}, -\frac{1}{2}, \frac{1}{2}, - \frac{1}{2} >~~, 
\end{equation}
and
\begin{equation}
\vert \downarrow >  = \vert 
-\frac{1}{2}, \frac{1}{2},- \frac{1}{2},  \frac{1}{2}, 
-\frac{1}{2}, \frac{1}{2}, -\frac{1}{2},  \frac{1}{2} >~~,
\end{equation}
and are eigenstates of Z-component of the Neel operator
$\vec{N} = \sum_{i=1,4}(\vec{\bf I}_{2i-1}-\vec{\bf I}_{2i})/4$ with
eigenvalues $+1$ and $-1$. \\ \indent
The results will be presented in detail for the particular case
of $d=7$ \AA, when $J= 0.17 \hbar/$ms and the anisotropy coefficients
given by Eq. (3) are $c_x=-0.22$, $c_y=-0.58$ and $c_z=1.2$.
The overlap coefficients between the eigenstates 
$\vert \psi_0>$ and $\vert \psi_1>$ and
the AF states of  Eq. (5) and (6) are  $< \uparrow \vert \psi_0> 
= < \downarrow \vert \psi_0 >= 0.57$, $- < \uparrow \vert \psi_1> 
= <  \downarrow \vert \psi_1 > = 0.63$. Thus, the two
AF states exhaust more than 65 \% of the eigenstates norm.
The eigenvalues $E_0=-0.406 \hbar/$ms and $E_1=-0.395 \hbar/$ms are
separated by $\Delta=  E_1 - E_0= 0.011 \hbar/ $ms  which is
sensibly smaller than $E_2-E_1 = 0.053 \hbar/$ms. 
The small value of $\Delta$ shows that the system has a quasi-degenerate
ground state.  This appears as a
"tunneling doublet" determined by the two-dimensional potential barrier
$E^A_{max} ( \phi ) = (0.066 \cos^2 \phi + 0.17 \sin^2 \phi) \hbar/ $ms
separating the AF energy minima of $E^A_{min}=-0.353 \hbar/$ms. \\ \indent
The tunneling behavior is shown clearly by the evolution
of the non-stationary wave-packets prepared at the ground-state energy 
with a well-defined AF configuration.
States with these properties are represented by the linear combinations
\begin{equation}
\vert \psi_{\downarrow}> = \frac{1}{ \sqrt{2}} ( 
\vert \psi_0 >+ \vert \psi_1>) ~~,~~
\vert \psi_{\uparrow} >= \frac{1}{ \sqrt{2}} ( 
\vert \psi_0 > - \vert \psi_1>) ~~.
\end{equation}
of the eigenstates $\vert \psi_0 >$ and $\vert \psi_1>$.
These wave-packets can be distinguished macroscopically by the
expectation value of $N_z$, and during the time-evolution are interchanged
periodically, by MQC oscillations. Thus, if the  system is prepared
at t=0 in the state $\vert \psi_\downarrow >$, then at the moment t it
will be found in the state $\vert  \psi_\uparrow > $ with the probability
${\cal P}_\uparrow (t)= \sin^2(\pi t/2T_{max})$, where 
$T_{max}= \hbar \pi / \Delta $. The half-period of oscillation $T_{max}$
is represented as a function of the chain constant
$d$ in Fig. 1 (A). For $d=7$ \AA, one obtains  $T_{max}= 0.3$ s, and
the calculation of the expectation values 
\begin{equation}
<N_k>(t)=< \psi_\downarrow \vert e^{i H_0 t/ \hbar} N_k e^{-i H_0t / \hbar}
\vert \psi_\downarrow >
\end{equation}
shows that in time $<N_x>(t)=0$, $<N_y>(t)=0$, while the Z-component 
has MQC oscillations  $<N_z>(t)=
< \psi_0 \vert N_z \vert \psi_1 > \cos(\pi t/T_{max})$, 
with $< \psi_0 \vert N_z \vert \psi_1 >= -0.83$
(Fig. 1 (B)). \\ \indent
It is important to emphasize the extreme sensitivity of the
MQC resonance with respect to the preparation of the initial state.
The expected value of the energy in the classical antiferromagnetic
states, $< \downarrow  \vert H_0 \vert \downarrow >$ and  $< \uparrow \vert
H_0 \vert \uparrow >$, is $E^A_{min}=-  0.353 \hbar/$ms,  higher than 
$E_1= - 0.39 \hbar/$ms by $\approx 4 \Delta$. This energy
is far outside the interval $[E_0,E_1]$,
and therefore between the classical antiferromagnetic states
$\vert \downarrow>$ and $\vert \uparrow >$  there are no
MQC oscillations. \\[.5cm] 
{\bf III. MQC oscillations at finite temperature}
\\[.5cm] \indent
The evolution of the Neel vector at the MQC resonance, presented in the
previous section, was obtained neglecting the residual interactions between
the nuclear spins  and the environment. However, when these interactions
are considered the  MQC oscillations could be damped \cite{bend} or
completely suppressed \cite{bm}, and the Neel vector aquires a fixed
orientation.
At finite temperatures the nuclear spin $\vec{\bf I}_i$  of each Xe atom
lying on the magnetic surface is subject also to an external, time-dependent 
magnetic field $\vec{B}_i^e(t)$, created by  
the phonon modulation of the  crystalline electric field and the 
lattice spin waves  \cite{kittel}. Due to this
field, appears in the Hamiltonian a residual interaction term 
\begin{equation}
H_r(t) =-  \gamma \hbar \sum_{i=1,8} \vec{\bf I}_i \cdot \vec{B}_i^e (t)~~.
\end{equation}
At low temperatures the typical wave length of the surface
phonons and magnons \cite{cottam} is  $\sim 400$ \AA,
much larger than the length  of the chain, $L= 49 $ \AA.
Therefore $\vec{B}_i^e$ will be considered to be the same for all spins, 
$\vec{B}^e_i(t) \equiv \vec{B}^e(t)$. With these approximations
the residual interaction term determined by the environmental magnetic field
is
$H_r (t)=-  \gamma \hbar \vec{B}^e (t) \cdot \vec{\bf I} $, where
$\vec{\bf I}= \sum_{i=1,8} \vec{\bf I}_i $. \\ \indent
The "atomic switch" experiments \cite{eig} have been performed at the
environmental temperature $T=4$K, when the thermal energy $k_B T = 0.34$ meV
is very high compared both to the tunnel splitting  $\Delta= 6.83 \cdot
10^{-12} $ meV and to the maximum barrier height
$V_B=E^A_{max} ( \pi/2) = 0.11 \cdot 10^{-9}$ meV.
Therefore, the thermal environment can be considered as classical, and the 
field components  $B^e_{\mu} (t)$, $\mu=x,y,z$, will be treated as a white
noise with zero mean. The normalization of this fluctuating field 
is ensured by the fluctuation-dissipation
theorem (FDT) $<< B^e_{\mu} (t) B^e_{\mu '} (t' ) >> = \delta_{\mu , \mu '}
\delta(t-t')/( \gamma^2 \tau )$,
where  $<<...>>$ denotes the average over the statistical ensemble describing
the environment. \\ \indent
For a single spin of the chain, $H_r(t)$ induces
transitions between the states $\vert 1/2> $ and 
$ \vert - 1/2> $ with a rate \cite{mg4}  $\lambda =
(\vert <-1/2 \vert  {\bf I}_x \vert 1/2 > \vert^2 +
\vert <-1/2 \vert  {\bf I}_y \vert 1/2 > \vert^2)/ \tau = 1/ 2 \tau$,
and the relaxation rate of the population difference 
$n_{1/2} - n_{-1/2}$ is $2 \lambda = 1/ \tau$. Therefore, the parameter
$\tau$ has the meaning of spin-surface relaxation time, 
increasing as $1/T$ when $T \rightarrow 0$. 
\\ \indent
The relaxation of the MQC oscillations cannot be treated by using a similar
two-level approximation, because the matrix elements of 
$H_r$ within the subspace generated by $\vert \psi_0>$ and
$\vert \psi_1 >$ are 0. The operators ${\bf I}_x$ and ${\bf I}_y$ contained
in $H_r$  act on the initial
state  $\vert \psi_\downarrow >$  by flipping the individual
spins, and in time the quantum state aquires components over the
whole spectrum. This process is described in principle by a transport
equation for the density matrix \cite{zurek, cl} but due to the relatively
large number of states ($=2^8$), such numerical calculations are not feasible.   
Moreover, at high temperatures, when the initial state of the system
is a pure state and there are only few observable of interest, as in the
present case, the computational effort required to find the evolution of the
whole density matrix is not justified. Instead, an equivalent
description \cite{mg4} which can be applied efficiently is provided by a
statistical ensemble of $N_t$ Brownian trajectories  $\vert \psi^r (t)> $,
$r=1,N_t$, obtained by integrating the Schr\"odinger equation
\begin{equation}
i \hbar \partial_t \vert  \psi^r(t)> = \lbrack H_0 - \gamma \hbar 
\vec{B}^e (t) \cdot \vec{\bf I} \rbrack  \vert \psi^r(t)>~~.
\end{equation} 
The numerical integration was performed using the procedure presented in
\cite{mg1}, as a classical system of Hamilton equations for the
real and immaginary  parts of the 
amplitudes $y_k(t)= < k \vert \psi (t)>$, $k=1,2^8$. Using the notation
$u_k (t) \equiv Re ( y_k(t) )$  and $v_k (t) \equiv Im ( y_k(t ))$,
Eq. (10) takes the form 
\begin{equation}
2 \hbar \dot{u}_k =  \frac{ \partial {\cal H}(t) }{ \partial v_k} ~~~~~~~~~~
2 \hbar \dot{v}_k = - \frac{ \partial {\cal H}(t) }{\partial u_k}~~,
\end{equation}
where
$$
{\cal H}(t) = \sum_{k,k'=1}^N (u_ku_{k'}+ v_k v_{k'} )   
Re(<k \vert H_0 +H_r(t)  \vert k'>) -
$$
\begin{equation}
(u_k v_{k'}-v_k u_{k'}) Im(<k \vert H_0 + H_r(t)  \vert k'>)~~.
\end{equation}
The dominant AF states are annihilated by ${\bf I}_z$, and therefore
the contribution of the term $B^e_z (t) {\bf I}_z$ from  $H_r$
is very small, and it was neglected. The remaining X and Y components of the
fluctuating field at the moment $t_n=ndt$, normalized according to
the FDT,  have the form
$
B^e (t_n) = R_n   \sqrt{  1 /( \gamma^2 \tau dt)} 
$
where $\{ R_n$, $n=1,2,3,.... \}$  is a sequence of random numbers 
with $0$  mean and variance 1.
\\ \indent
The time-evolution of the ensemble average for any observable ${\cal O}$ is
defined by
\begin{equation}
<< {\cal O} >>(t)= \frac{1}{N_t} \sum_{r=1}^{N_t} 
< \psi^r (t) \vert {\cal O} \vert \psi^r (t)>~~. 
\end{equation}
The observables of interest here are the Neel vector and
the energy, and their averages were calculated using Eq. (13) 
with ${\cal O}$ denoting the components of the Neel operator $N_k$, and
$H_0$, respectively.  The results obtained for $d=7$ \AA, $\tau=2.5$ s and
$N_t=20$ are presented in Fig. (2). 
The components $<<N_x>>$ and $<<N_y>>$ of Fig. 2 (A) and (B) have thermal
fluctuations around 0, while $<<N_z>>$ shown in Fig. 2 (C)
has damped oscillations. These oscillations can be well approximated by
the analytical expression
\begin{equation}
<<N_z>> (t) =  e^{- \Lambda t} <N_z>(t)~~,
\end{equation}
where the damping constant obtained by fit is $\Lambda = 3 $ s$^{-1}$.
Similar calculations with $\tau$ in the range of seconds, as
expected at low temperatures, indicate that $\Lambda= 7.5 / \tau$.
\\ \indent
The ratio $(<<H_0>>(t)-E_0)/ \Delta$
between the average excitation energy and the doublet splitting $\Delta$ is a
suitable measure of the  heating effect produced by the environment, and
is represented in Fig. 2 (D). The ensemble average of the energy 
appearing here is accurately reproduced by the formula
$<<H_0>>(t)= <<H_0>>(0)+ \Delta( w_1 t- w_2 t^2 )$, with
$w_1= 45.9 $ s$^{-1}$ and $w_2=19.3$ s$^{-2}$. 
\\[.5cm] 
{\bf  IV. Conclusions} \\[.5cm]
In this work it was shown that
the chain of nuclear spins for eight atoms of the $^{129}$Xe
isotope placed on a magnetic surface has a low-energy mode of collective
excitation represented by antiferromagnetic MQC oscillations.
The numerical calculations indicate that when the  chain
constant $d$ is larger than $6.8$ \AA$~~$ the first excited state $\psi_1$
of the chain has a very low excitation energy.
This state is strongly coupled to the ground state $\psi_0$ by the
Z-component of the Neel operator,  such that
$\vert < \psi_0 \vert N_z \vert \psi_1 > \vert= 0.83$. 
Both states, $\psi_0$ and $\psi_1$, are dominated by two AF spin
configurations having opposite orientations of the Neel vector.  Therefore,
resonant wave packets containing only one of these AF configurations
can be constructed by the symmetric and antisymmetric linear combinations
of $\psi_0$ and $\psi_1$. These wave-packets are non-stationary, and are
interchanged periodically in time.
Although the number of spins considered here is relatively small, 
the resonant wave packets are distinguished by 
the orientation of the Neel vector,  and their oscillation represent a MQC
phenomenon.
When $d=7$ \AA$~~$ the half-period of oscillation is $T_{max}=0.3$ s, and
increases exponentially with $d$ (Fig. 1 (A)).  \\ \indent
The MQC oscillations are known to be very sensitive to the decoherence
effects produced by the coupling to the thermal environment. 
If in practice these effects are strong, then the oscillations may be
completely suppressed and cannot be observed.  
In the present case decoherence appears due to the residual dipole interaction
between the nuclear spins and the fluctuating magnetic field produced by the
surface. This coupling was described by one parameter, chosen to be 
the relaxation time  $\tau$ of a single nuclear spin. \\ \indent
The relaxation rate of the MQC oscillations can be easily calculated in the
physical situations which can be treated within a two-level approximation.  
However, for eight spins the number of states coupled by the residual dipole 
interaction term is large, and this approximation cannot be
applied. Therefore, the fluctuating residual interaction term was included
in the Hamiltonian, and the diffusion of the initial wave-packet
$\vert \psi_\downarrow >$  was described by an ensemble of Brownian solutions
obtained by integrating the Schr\"odinger equation. The results indicate
that when $\tau$ is in the range of seconds, the MQC oscillations are not
suppressed, but are damped by a rate $ \sim (N-1)/ \tau$, where $N$ is the
number of spins.     \\

{\em ACKNOWLEDGMENTS: }One of the authors (MS) is thankful to NSERC of
Canada for financial support in the form of a research grant.

{\bf Figure Captions} \\[.5cm]

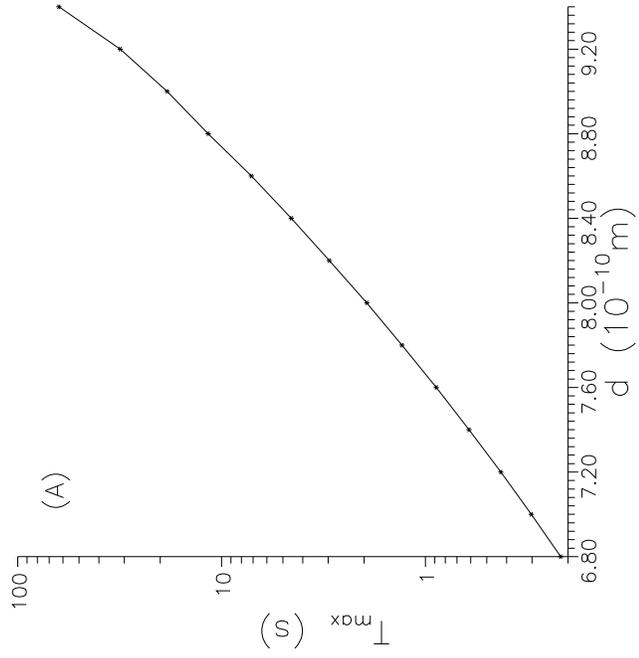
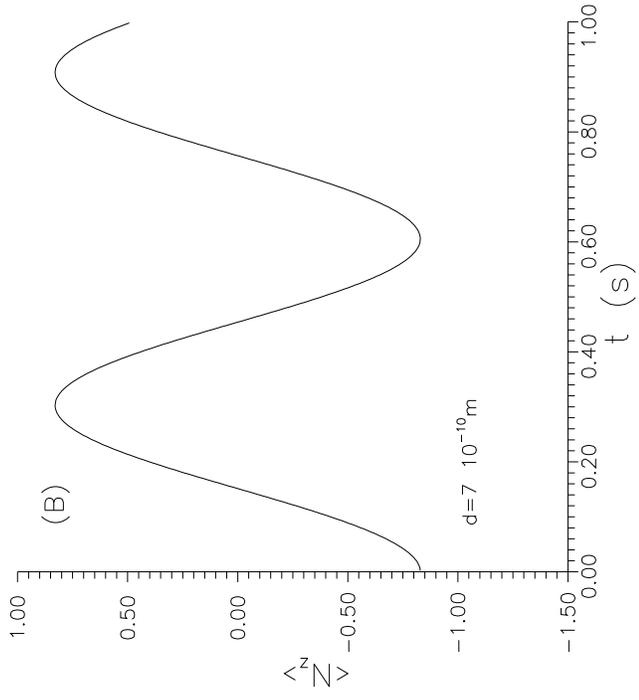
Fig. 1. The half-period of the MQC oscillations $T_{max}$
as a function of the chain constant $d$
(A) and the expected value $< N_z>$ as a function of time 
when $d=7$ \AA$~~$ (B).    \\

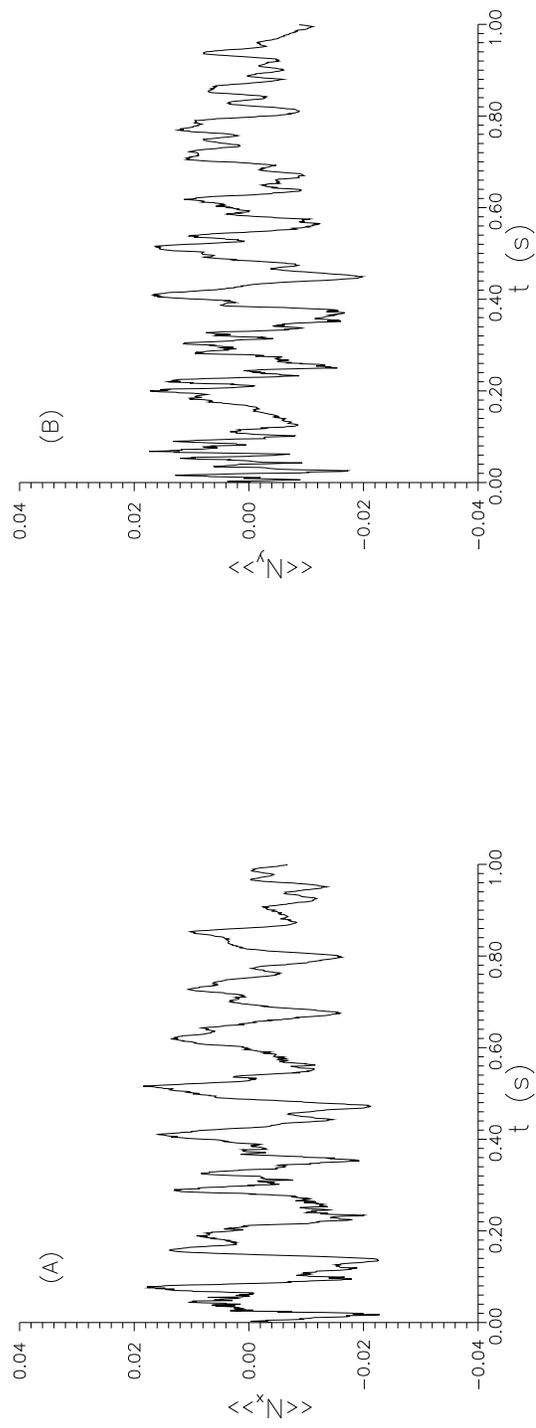
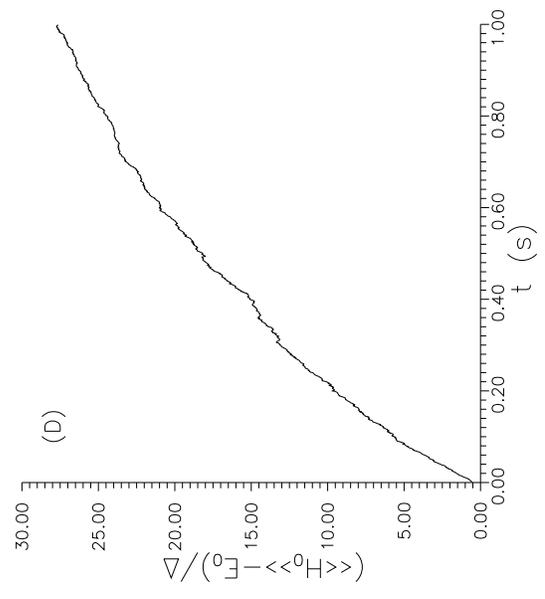
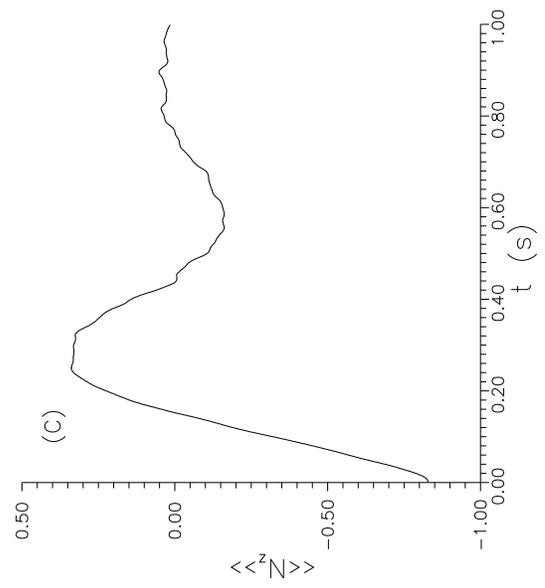
Fig. 2. Ensemble averages of the Neel vector components $<<N_x>>$ (A),
$<<N_y>>$ (B), $<<N_z>>$ (C) and
$(<<H_0>>-E_0)/ \Delta$ (D) as functions of time when
$d=7$ \AA$~~$ and $\tau=2.5$ s.

\end{document}